\documentclass[pra,aps,reprint,superscriptaddress,amsmath]{revtex4-1}

\usepackage{bm}
\usepackage{graphicx}
\usepackage{dsfont}
\usepackage{xcolor}

\newcommand{\bra}[1]{\langle #1 | \,}
\newcommand{\ket}[1]{\, | #1 \rangle}
\newcommand{\braket}[2]{\langle #1 | #2 \rangle}
\newcommand{\expv}[1]{\langle #1 \rangle}

\newcommand{\mr}[1]{\mathrm{#1}}
\newcommand{\mc}[1]{\mathcal{#1}}

\newcommand{\om}{\omega}

\newcommand{\eps}{\epsilon}
\newcommand{\veps}{\varepsilon}

\begin{document}

\title{Collective emission of photons from dense, dipole-dipole interacting atomic ensembles}

\author{David Petrosyan}
\affiliation{Institute of Electronic Structure and Laser,
Foundation for Research and Technology -- Hellas, GR-70013 Heraklion, Crete, Greece}
\affiliation{Department of Physics and Astronomy, Aarhus University, DK-8000 Aarhus C, Denmark}

\author{Klaus M\o lmer}
\affiliation{Department of Physics and Astronomy, Aarhus University, DK-8000 Aarhus C, Denmark}

\date{\today}

\begin{abstract}
We study the collective radiation properties of cold, trapped ensembles of atoms.
We consider the high density regime with the mean interatomic distance being comparable to,
or smaller than, the wavelength of the resonant optical radiation emitted by the atoms.
We find that the emission rate of a photon from an excited atomic ensemble
is strongly enhanced for an elongated cloud. We analyze collective single-excitation
eigenstates of the atomic ensemble and find that the absorption/emission spectrum
is broadened and shifted to lower frequencies as compared to the non-interacting (low density)
or single atom spectrum. We also analyze the spatial and temporal profile of the emitted
radiation. Finally, we explore how to efficiently excite the collective super-radiant
states of the atomic ensemble from a long-lived storage state in order to implement
matter-light interfaces for quantum computation and communication applications.
\end{abstract}

\maketitle


\section{Introduction}

Super- and sub-radiance have been an active topic of research since the
seminal paper of Dicke \cite{Dicke1954} on collective emission of atoms confined
within a distance that is small compared to the wavelength of the resonantly emitted radiation.
An extended atomic ensemble prepared by a laser field in the excited ``timed Dicke'' state
collectively decays by an emission of a photon predominantly into the phase-matched direction
\cite{Scully2006,Mazetz2007}, while multiple scattering and reabsorption of photons in
large atomic clouds modifies the exponential decay of the collective atomic excitation
\cite{Mazetz2007,SvidzinskyScully2008,Bienaime2011}. The behavior of the single excitation
states of the atoms can be understood in terms of the collective eigenstates of an effective non-Hermitian Hamiltonian
\cite{Svidzinsky2008,Svidzinsky2010} which exhibit enhanced (super-radiant) and
suppressed (sub-radiant) decay rates, together with level shifts (collective Lamb shift).
Recent experiments have demonstrated both sub-radiance \cite{Bienaime2012,Guerin2016}
and super-radiance \cite{Oliveira2014,Ortiz-Gutierrez2018,Araujo2016,Roof2016,Guerin2019} in large, dilute atomic clouds.
In this article, we calculate the spectral and spatio-temporal properties of the emitted radiation
for various trapping geometries of the atomic clouds, taking into account the interatomic interactions
in the high atom density regime \cite{Lehmberg1970,Thirunamachandran,Jenkins2016,Corman2017}.
We also consider the excitation of the collective super-radiant states from a long-lived storage state 
of atoms for Raman conversion of an atomic spin-wave into an optical photon.

In addition to the fundamental interest in the physical processes, our study is motivated
by practical applications of matter-light interfaces for quantum information processing and communications.
Atomic ensembles have good coherence properties and strong dipole transitions for
efficient coupling to optical photons \cite{EITrev2005,Hammerer2010,Sangouard2011}.
Moreover, atoms can couple to microwave fields and thereby be interfaced with superconducting circuits,
which are currently among the most advanced candidates for quantum processors \cite{Kurizki2015}.
The atomic ensembles can then play the role of quantum memories and microwave to optical
transducers \cite{Kurizki2015,Petrosyan2019,Covey2019,Lauk2020}. In turn, photons can serve as flying
qubits to encode and reliably transmit quantum information over long-distances \cite{Kimble2008,OBrien2009}.
For optical photons, transmission may occur through free space or via fiber waveguides, and it is important
to determine the spatio-temporal profile of the photon emitted by the atoms to optimally construct
the paraxial optical elements that will collect the photon and direct it to a distant receiver \cite{Kurko}.

This paper is structured as follows.
In Sec. \ref{sec:formalism}, we present the mathematical formalism to describe the quantum interactions
between $N$ cold atoms at random positions in a trap and the quantized radiation field mediating interatomic
interactions and their collective emission.
In Sec. \ref{sec:twoLAs}, we discuss solutions for two-level atoms and identify how the trapping geometry
influences the emission properties of the system.
In Sec. \ref{sec:threeLAs}, we extend the analysis to three-level atoms where the initial excitation is driven 
from a third storage state, and the excitation dynamics is influenced by the atomic interactions.
Section~\ref{sec:concl} concludes the article and discusses the prospects of applications.

\section{Quantum interactions between atoms and light}
\label{sec:formalism}

\begin{figure}[t]
\centering
\includegraphics[width=0.9\linewidth]{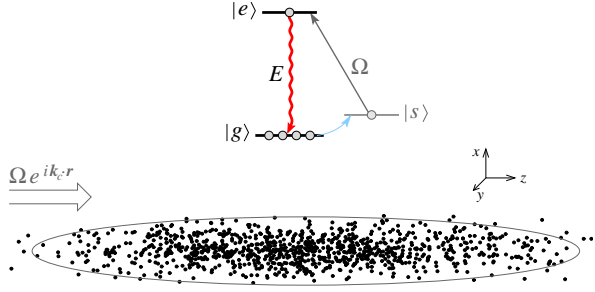}
\caption{Internal states and spatial configuration of an ensemble of cold atoms
with random positions in an elongated harmonic trap.
With all the atoms initially in the ground state $\ket{g}$, a weak microwave
or Raman transition (light-blue arrow) creates a single collective excitation in the
storage state $\ket{s}$. The atoms can be transferred from the storage state to the excited
state $\ket{e}$ by a laser pulse with Rabi frequency $\Omega$ and wavevector $\bm{k}_c$.
The excited state atoms decay to the ground state by emitting a photon into the free-space
radiation field $E$.}
\label{fig:Scheme}
\end{figure}

We consider $N \gg 1$ cold atoms at random positions in a harmonic trap, with the density distribution
\begin{equation}
\rho(\bm{r}) = N \frac{\exp \left( - \frac{x^2}{2\sigma_x^2} - \frac{y^2}{2\sigma_y^2} - \frac{z^2}{2\sigma_z^2} \right)}
{(2\pi)^{3/2} \sigma_x \sigma_y \sigma_z} . \label{eq:Gaussdens}
\end{equation}
The relevant internal states of the atoms are the ground state $\ket{g}$, an electronically excited state $\ket{e}$,
and a long-lived storage state $\ket{s}$, as shown in Fig.~\ref{fig:Scheme}.

We denote the collective ground state of the atoms as $\ket{G} \equiv \ket{g_1,g_2,\ldots,g_N}$.
A weak (single-photon) microwave or Raman process that acts symmetrically on all the atoms
can transfer the ground state to the collective single-excitation storage state
$\ket{S} = \frac{1}{\sqrt{N}} \sum_{j=1}^N \ket{g_1,g_2,\ldots,s_j, \ldots, g_N}$.
Subsequently, a spatially uniform laser pulse can near-resonantly couple the storage state $\ket{s}$
to the excited state $\ket{e}$ with Rabi frequency $\Omega$. An atom in the excited state $\ket{e}$
can decay to the ground state $\ket{g}$ by emitting a photon into the free-space radiation field
$\hat{\bm{E}}(\bm{r}) = \sum_k \hat{a}_k \bm{u}_k(\bm{r})$, where $\hat{a}_k$ are the bosonic annihilation
operators for the plane-wave modes
$\bm{u}_k(\bm{r}) = \hat{\bm{\veps}}_{\bm{k},\sigma} \sqrt{\frac{\hbar \omega_k}{2 \eps_0 V}} e^{i \bm{k} \cdot \bm{r}}$
forming a complete basis for the field within the quantization volume $V$.
The Hamiltonian of the system is
\begin{eqnarray}
H &=& \sum_k \hbar \omega_k \hat{a}_k^{\dag} \hat{a}_k + \sum_{j=1}^N \sum_{\mu =g,s,e} \hbar \omega_{\mu} \ket{\mu}_j\bra{\mu}
\nonumber \\ & &
- \sum_{j=1}^N [\hbar \Omega e^{i (\bm{k}_c \cdot \bm{r}_j - \omega_c t)} \ket{e}_j\bra{s}
\nonumber \\ & & \qquad \quad
+ \bm{\wp}_{eg} \cdot \hat{\bm{E}}(\bm{r}_j) \ket{e}_j\bra{g} + \mathrm{H. c}] ,
\end{eqnarray}
where the first term on the r.h.s. is the Hamiltonian for the field modes with energies $\hbar \omega_k$,
the second term corresponds to the Bohr energies $\hbar \omega_{\mu}$ of the atomic levels $\ket{\mu}$ ($\mu = g,s,e$),
the third term describes the interaction of the atoms at positions $\bm{r}_j$ with the coupling laser with frequency
$\omega_c$ and wavevector $\bm{k}_c \parallel \hat{\bm{z}}$, and the last term describes the coupling of the atoms
to the quantized free-space radiation field with the dipole moment $\bm{\wp}_{eg}$ on the transition $\ket{e} \to \ket{g}$.
For simplicity, we neglect the coupling of the atoms with the free-space radiation
and the resulting decay on the $\ket{e} \to \ket{s}$ transition \cite{NOgsdecay}.
We set the energy of the ground state to zero, $\hbar \omega_g=0$, and assume that
$\omega_s \ll \omega_e$ ($\omega_c \simeq \omega_e$).

The state vector of the system with the single atomic or photonic excitation can be expanded as
\begin{eqnarray*}
\ket{\Psi} &=& \sum_j c_j e^{-i \omega_s t} \ket{s_j} \otimes \ket{0} + \sum_j b_j e^{-i \omega_e t} \ket{e_j}\otimes \ket{0}
\nonumber \\ & &
+ \ket{G} \otimes \sum_k a_k e^{-i \omega_k t} \ket{1_k} ,
\end{eqnarray*}
where $\ket{G} \equiv \ket{g_1,g_2,\ldots,g_N}$,
$\ket{s_j} \equiv \ket{g_1,g_2,\ldots,s_j, \ldots, g_N}$,
$\ket{e_j} \equiv \ket{g_1,g_2,\ldots,e_j, \ldots, g_N}$,
while $\ket{1_k} \equiv \hat{a}_k^{\dag}\ket{0}$ denotes the state
of the radiation field with a single photon in mode $k$.
The state vector evolves according to the Schr\"odinger equation
$\partial_t \ket{\Psi} = -\frac{i}{\hbar} H \ket{\Psi}$,
leading to a set of equations for the atomic amplitudes
\begin{subequations}
\label{eqs:cjbj}
\begin{eqnarray}
\partial_t c_j &=& i \Omega^* e^{-i\bm{k}_c \cdot \bm{r}_j } b_j e^{i \Delta_c t}, \label{eqs:cj} \\
\partial_t b_j &=& i \Omega  e^{i \bm{k}_c \cdot \bm{r}_j} c_j e^{- i \Delta_c t}  + i \sum_k g_k(\bm{r}_j) a_k e^{i(\omega_e-\omega_k)t} \quad
\label{eqs:bj}
\end{eqnarray}
\end{subequations}
with $\Delta_c = \omega_c - \omega_e$, and an equation for the field amplitudes cast in the integral form
\begin{equation}
a_k(t) = i \sum_j g_k^*(\bm{r}_j) \int_0^t dt' b_j(t') e^{i(\omega_k-\omega_e)t'} , \label{eq:ak}
\end{equation}
where $g_k(\bm{r}_j) = \frac{\bm{\wp}_{eg} \cdot \bm{u}_k(\bm{r}_j)}{\hbar}$ is the atom-field coupling strength.

\subsection{Atoms}

Let us for the moment disregard the storage state and the transition $\ket{s} \to \ket{e}$,
assuming $\Omega = 0$, and consider two-level atoms with the ground $\ket{g}$ and excited $\ket{e}$ states.
We substitute Eq.~(\ref{eq:ak}) into Eq.~(\ref{eqs:bj}) and use the Born-Markov approximation to
eliminate the radiation field \cite{Lehmberg1970,Thirunamachandran}, obtaining
a closed set of equations for the atomic amplitudes
\begin{equation}
\partial_t b_j = - \frac{1}{2} \Gamma \, b_j
- \frac{1}{2} \Gamma \sum_{i\neq j} F_{ji} b_i . \label{eqs:bjGF}
\end{equation}
Here $\Gamma = \frac{1}{4 \pi \eps_0} \frac{4 k_e^3 |\wp_{eg}|^2}{3 \hbar}$ is the usual spontaneous decay rate
of the atom in the excited state $\ket{e}$ whose Lamb shift can be incorporated into $\omega_e$
\cite{ScullyZubary1997,PLDP2007}, and $F_{ji} = f_{ji} + i g_{ji}$ is the complex dipole-dipole
exchange interaction (including the near-field terms) between the atoms,
\begin{eqnarray*}
f_{ji} & =& \frac{3}{2} \big[ 1 - (\hat{\bm{\wp}} \cdot \hat{\bm{r}}_{ij})^2 \big] \frac{\sin (k_e r_{ij})}{k_e r_{ij}}
\nonumber \\ & &
+ \frac{3}{2} \big[ 1 - 3 (\hat{\bm{\wp}} \cdot \hat{\bm{r}}_{ij})^2 \big]
\left[ \frac{\cos (k_e r_{ij})}{(k_e r_{ij})^2} - \frac{\sin (k_e r_{ij})}{(k_e r_{ij})^3} \right]  , \\
g_{ji} & =& - \frac{3}{2} \big[ 1 - (\hat{\bm{\wp}} \cdot \hat{\bm{r}}_{ij})^2 \big] \frac{\cos (k_e r_{ij})}{k_e r_{ij}}
\nonumber \\ & &
+ \frac{3}{2} \big[ 1 - 3 (\hat{\bm{\wp}} \cdot \hat{\bm{r}}_{ij})^2 \big]
\left[ \frac{\sin (k_e r_{ij})}{(k_e r_{ij})^2} + \frac{\cos (k_e r_{ij})}{(k_e r_{ij})^3} \right]  ,
\end{eqnarray*}
where $\hat{\bm{\wp}} \equiv \frac{\bm{\wp}_{eg}}{\wp_{eg}}$ is the unit vector in the direction of the atomic
dipole moment, $\hat{\bm{r}}_{ij} \equiv \frac{\bm{r}_{ij}}{r_{ij}}$ is the unit vector along the direction
of the relative position vector $\bm{r}_{ij} = \bm{r}_{i} -\bm{r}_{j}$ between atoms $i$ and $j$, $r_{ij} \equiv |\bm{r}_{ij}|$
is the distance between the atoms, and $k_e = \omega_e/c = 2\pi/\lambda_e$ with $\lambda_e$ being the wavelength of
the resonant photon. Note that for an isotropic dipole moment
$(\hat{\bm{\wp}} \cdot \hat{\bm{r}}_{ij})^2 = \frac{1}{3} \; \forall \; \hat{\bm{r}}_{ij}$
(or for sufficiently low density with the mean interatomic separation $\expv{r_{ij}} > \lambda_e$)
the interatomic interaction takes the simple form $F_{ji} = \frac{e^{i k_e r_{ij}}}{i k_e r_{ij}}$
(or $F_{ji} \simeq \frac{3}{2} \big[ 1 - (\hat{\bm{\wp}} \cdot \hat{\bm{r}}_{ij})^2 \big]\frac{e^{i k_e r_{ij}}}{i k_e r_{ij}}$)
\cite{Svidzinsky2010,Bienaime2011}.

\subsection{Radiation Field}
Consider now the wavefunction \cite{ScullyZubary1997,Miroshnychenko2013} of the emitted single-photon field
\begin{eqnarray}
\bm{E}(\bm{r},t) & \equiv & \bra{0} \bra{G} \hat{\bm{E}}(\bm{r}) \ket{\Psi (t)}
= \sum_k \bm{u}_k (\bm{r}) a_k(t) e^{-i \omega_k t}
\nonumber \\
&=& i \frac{\wp_{eg}}{2 \eps_0 V} \sum_{\bm{k},\sigma} \hat{\bm{\veps}}_{\bm{k},\sigma}
(\hat{\bm{\wp}} \cdot \hat{\bm{\veps}}_{\bm{k},\sigma}) \, \omega_k
\sum_j e^{i \bm{k}\cdot ( \bm{r} - \bm{r}_j)}
\nonumber \\ & & \qquad \times
\int_0^t dt' b_j(t') e^{-i \omega_k (t-t') - i \omega_e t'} .
\end{eqnarray}
We sum over the two orthogonal photon polarizations $\sigma=1,2$ for each $\bm{k}$
($\hat{\bm{\veps}}_{\bm{k},\sigma} \perp \bm{k}$), $\sum_\sigma \hat{\bm{\veps}}_{\bm{k},\sigma} \cdot \hat{\bm{\veps}}_{\bm{k},\sigma}
= \mathds{I} - \hat{\bm{k}} \otimes \hat{\bm{k}}$, where $\mathds{I}$ is the unity tensor and $\hat{\bm{k}} \equiv \frac{\bm{k}}{k}$,
and replace the summation over the modes $\bm{k}$ by an integration via
$\sum_{\bm{k}} \to \frac{V}{(2\pi)^3} \int d^3 k = \frac{V}{(2\pi)^3}  \int_0^{\infty} d k \, k^2 \int_{4\pi} d \Omega_k$
\cite{ScullyZubary1997,PLDP2007}, obtaining
\begin{eqnarray}
\bm{E}(\bm{r},t) &=& i \frac{\wp_{eg}}{2 (2\pi)^3 \eps_0} \sum_j  \int_0^t dt' b_j(t') e^{- i \omega_e t'}
\nonumber \\ & & \qquad \times
\int_0^{\infty} dk k^2 \omega_k  e^{-i \omega_k (t-t')}
\nonumber \\ & & \qquad \times
\int_{4\pi} d \Omega_k    e^{i \bm{k} \cdot (\bm{r} -\bm{r}_j) }
[\mathds{I} - \hat{\bm{k}} \otimes \hat{\bm{k}} ] \cdot \hat{\bm{\wp}},
\label{eq:Ertplw}
\end{eqnarray}
The integration over the $4\pi$ solid angle with $d \Omega_k = \sin \theta d \theta d \varphi$ leads to
$4\pi \frac{\sin (k |\bm{r} -\bm{r}_j|) }{k |\bm{r} -\bm{r}_j|} [\mathds{I} - \hat{\bm{r}}_j \otimes \hat{\bm{r}}_j]
= -i \frac{2\pi c [\mathds{I} - \hat{\bm{r}}_j \otimes \hat{\bm{r}}_j ]}{\omega_k|\bm{r} -\bm{r}_j|}
(e^{i k |\bm{r} -\bm{r}_j|} - \mathrm{c.c})$, where $\hat{\bm{r}}_j \equiv \frac{\bm{r}_j}{r_j}$.
We substitute this into the above equation, assume that during the photon emission $k$
is peaked around the atomic resonance $k_e = \omega_e/c$ and pull $k_e^2$ out of the integral,
and extend the lower limit of integration over $k$ to $-\infty$, as in the Weisskopf-Wigner
approximation \cite{ScullyZubary1997,Miroshnychenko2013}. We then have
\begin{eqnarray*}
&& \int_{-\infty}^{\infty} dk (e^{i k |\bm{r} -\bm{r}_j| -i ck (t-t')} - e^{-i k |\bm{r} -\bm{r}_j| -i ck (t-t')} )
\\ &&
=  \frac{2\pi}{c} \delta(t'-t + |\bm{r} -\bm{r}_j|/c) + \frac{2\pi}{c} \delta(t'-t - |\bm{r} -\bm{r}_j|/c) .
\end{eqnarray*}
Upon substitution into Eq.~(\ref{eq:Ertplw}) the second term is always zero, and we finally obtain
\begin{eqnarray}
\bm{E}(\bm{r},t) &=& \frac{\wp_{eg} k_e^2}{4\pi \eps_0}
\sum_j \frac{e^{-i \om_e (t-|\bm{r} -\bm{r}_j|/c)}}{|\bm{r} -\bm{r}_j|} b_j(t - |\bm{r} -\bm{r}_j|/c)
\nonumber \\ & & \qquad \qquad \times
[\mathds{I} - \hat{\bm{r}}_j \otimes \hat{\bm{r}}_j] \cdot \hat{\bm{\wp}} .
\end{eqnarray}

For a single atom at the origin, we have a (in general anisotropic) spherical wave
$\bm{E}(\bm{r},t) = \frac{\wp_{eg} k_e^2}{4\pi \eps_0} \frac{e^{-i \om_e (t- r/c)}}{r} b(t-r/c)
\, [\mathds{I} - \hat{\bm{r}}_j \otimes \hat{\bm{r}}_j] \cdot \hat{\bm{\wp}}$,
while the intensity of the emitted radiation in the direction of $\bm{r}$ is
$I_{\sigma}(\bm{r},t) = \frac{\eps_0 c}{2} |\hat{\bm{\veps}}_{\bm{r},\sigma} \cdot \bm{E}(\bm{r},t)|^2
= \frac{\hbar \omega_e}{4 \pi r^2} \frac{3 |\hat{\bm{\veps}}_{\bm{r},\sigma} \cdot \hat {\bm{\wp}}|^2}{8} \Gamma |b(t-r/c)|^2$
for each polarization component $\hat{\bm{\veps}}_{\bm{r},\sigma} \perp \bm{r}$. As an example, for $\Delta M = \pm 1$
atomic transition with $\hat {\bm{\wp}} = \frac{1}{\sqrt{2}} (\hat{\bm{x}} \pm i \hat{\bm{y}})$ we obtain
the dipole emission pattern $I_{1} + I_{2} \propto \frac{1}{2}(1+\cos^2 \theta)$.

In the far-field region, we have
$|\bm{r} -\bm{r}_j| \simeq r - (\bm{r} \cdot \bm{r}_j)/r = r - \hat{\bm{r}} \cdot \bm{r}_j$, and therefore
\begin{equation}
E_{\sigma}^{(\mathrm{ff})}(\bm{r},t) = (\hat{\bm{\veps}}_{\bm{r},\sigma} \cdot \hat{\bm{\wp}})
\frac{\wp_{eg} k_e^2}{4\pi \eps_0}  \frac{e^{i (k_er - \om_e t)}}{r}
\sum_j  b_j(t - r/c) e^{-i \bm{k}_e \cdot \bm{r}_j} , \label{eq:Erff}
\end{equation}
where $\bm{k}_e \equiv k_e \hat{\bm{r}}$.

\subsubsection*{Non-interacting atoms}
Consider an ensemble of $N$ atoms with density $\rho(\bm{r})$, such that $\int d^3 r \rho(\bm{r}) = N$,
prepared initially in the collective single-excitation (timed-Dicke \cite{Scully2006}) state
$\ket{E_{\mr{TD}}} = \frac{1}{\sqrt{N}} \sum_{j=1}^N e^{i \bm{k}_c \cdot \bm{r}_j} \ket{e_j}$.
For non-interacting atoms, $F_{ji} = 0$, we have from Eq.~(\ref{eqs:bjGF})
that $b_j(t) = \frac{1}{\sqrt{N}} e^{-\frac{1}{2} \Gamma t} e^{i \bm{k}_c \cdot \bm{r}_j}$.
Disregarding the photon polarization, the emitted field in the far-field region is
\begin{subequations}
\label{eqs:Erffini}
\begin{eqnarray}
E^{(\mathrm{ff})}(\bm{r},t) &=&
\frac{\wp_{eg} k_e^2}{4\pi \eps_0} \frac{e^{-\frac{1}{2} \Gamma (t-r/c)}}{\sqrt{N}} e^{- i \om_e t} \, \mc{E}(\bm{r}),  \\
\mc{E}(\bm{r}) &\equiv & \frac{e^{i k_e r }}{r} \sum_j e^{i (\bm{k}_c - \bm{k}_e) \cdot \bm{r}_j}
\nonumber \\
&=& \frac{e^{i k_e r }}{r} \int d^3 r' \rho(\bm{r}') e^{i (\bm{k}_c - \bm{k}_e) \cdot \bm{r}'} .
\end{eqnarray}
\end{subequations}
Substituting here the Gaussian density distribution $\rho(\bm{r}')$ of Eq.~(\ref{eq:Gaussdens}) and performing
the integration over $\bm{r}'$, we obtain
\begin{equation}
\mc{E}(\bm{r}) =
N \frac{e^{i k_e r }}{r} \exp \left \{ - \frac{k_e^2}{2} \left[ \frac{x^2 + y^2}{r^2} \sigma_{\perp}^2
+ \frac{(z - r)^2}{r^2} \sigma_z^2 \right] \right\} , \label{eq:ErffInt}
\end{equation}
where we assume that $\sigma_{x,y} = \sigma_{\perp}$. Consider the field amplitude $\mc{E}(\bm{r})$
along the $z$ direction within a small axial distance $\varrho = \sqrt{x^2 +y^2} \ll z$, such that
$r = \sqrt{z^2 + \varrho^2} \simeq z+ \frac{\varrho^2}{2z}$.
With $r^2 \approx z^2$ and $(r - z)^2 \approx 0$, we have from Eq.~(\ref{eq:ErffInt})
\begin{equation}
\mc{E}(\bm{r}) \approx \frac{N}{z +\frac{\varrho^2}{2z}}
\exp \left[ i k_e \left( z   + \frac{\varrho^2}{2z} \right) - \frac{k_e^2}{2} \frac{\varrho^2}{z^2} \sigma_{\perp}^2 \right] .
\label{eq:Erffapprx}
\end{equation}
On the other hand, a Gaussian field mode with the waist $w_0$ at $z=0$ has the form
\begin{equation}
\phi_{k}(\bm{r}) = \frac{\zeta_k}{q_k^*(z)} \exp \left[ i k \left( z + \frac{\varrho^2}{2q_k^*(z)} \right) \right] ,
\label{eq:Gfk}
\end{equation}
where $\zeta_k = k w_0^2/2$ is the Rayleigh length and $q_k(z) = z + i \zeta_k$ is the complex beam parameter.
In the far field, $z^2 + \zeta_k^2 \approx z^2$, we have
\begin{eqnarray}
\phi_{k}(\bm{r}) \approx \frac{\zeta_k}{z - i\zeta_k}
 \exp \left[ i k \left( z   + \frac{\varrho^2}{2z} \right) - \frac{k^2}{2} \frac{\varrho^2}{z^2} \frac{\zeta_k}{k} \right] .
\label{eq:Gfkapprx}
\end{eqnarray}
Comparing this with Eq.~(\ref{eq:Erffapprx}), we see that, apart from the Gouy phase that originates from
the imaginary part of $\frac{\zeta_k}{z - i\zeta_k}$, the far field $\mc{E}(\bm{r})$ is mostly emitted into
a Gaussian mode with wavevector $k=k_e$ and a beam waist determined from $\zeta_k/k = w_0^2/2 \approx \sigma_{\perp}^2$,
i.e., $w_0 = \sqrt{2} \sigma_{\perp}$, while the angular spread (divergence) of the beam is
$\Delta \theta = \frac{\lambda_e}{\pi w_0} = \frac{\sqrt{2}}{k_e \sigma_{\perp}}$.
More qualitatively \cite{Saffman2005,Petrosyan2018}, the probability of the cooperative
photon emission into the phase-matched direction within the solid angle
$\Delta \Omega = \pi (\Delta \theta)^2 = \frac{2 \pi}{(k_e \sigma_{\perp})^2}$,
as opposed to spontaneous, uncorrelated photon emission into the $4\pi$ solid angle, is
\begin{equation}
P_{\Delta \Omega} \simeq \frac{N \Delta \Omega}{4\pi + N \Delta \Omega} . \label{eq:PDeOm-nia}
\end{equation}

\section{Two-level atomic medium}
\label{sec:twoLAs}

\subsection{Collective decay dynamics of interacting atoms}

\begin{figure}[t]
\includegraphics[width=1.0\linewidth]{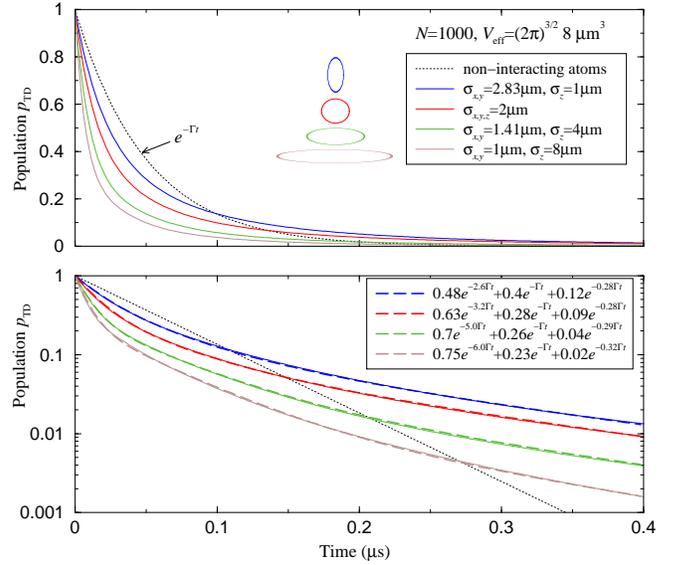}
\caption{Dynamics of the population $p_{\mathrm{TD}}(t) = |\braket{E_{\mr{TD}}}{\Psi(t)}|^2$
of the initially prepared timed-Dicke state $\ket{E_{\mr{TD}}}$ of the atoms in a harmonic 
trap with different aspect ratios $\sigma_{x,y} / \sigma_z$; The progressively lower curves 
correspond to decreasing $\sigma_{x,y}$ and increasing $\sigma_z$, with the product 
$\sigma_x \sigma_y \sigma_z$ kept constant. We place $N = 1000$ atoms at random positions in
an effective volume $V_{\mathrm{eff}} = (2 \pi)^{3/2} \sigma_x \sigma_y \sigma_z = (2 \pi)^{3/2} 8\,\mu\mathrm{m}^3$,
i.e., the mean interatomic separation $\expv{r_{ij}} = \sqrt[3]{V_{\mathrm{eff}}/N} \simeq 0.5\,\mu\mathrm{m}$.
Each curve corresponds to a single realization of the ensemble of atoms at random positions, but different realizations
for the same atom number and trap geometry give very similar results (for large enough $N$ as here).
The wavelength of the resonant transition $\ket{e} \to \ket{g}$ is $\lambda_e = 780\:$nm,
decay rate $\Gamma = 2 \times 10^7\:\mr{s}^{-1}$, and the transition dipole moment is along
$\hat{\bm{\wp}} = \frac{\hat{\bm{x}} + i \hat{\bm{y}}}{\sqrt{2}}$ ($\Delta M = 1$ transition).
In the lower panel, we fit the population decay curves with the sum of three exponential terms,
$p_{\mathrm{TD}}(t) \simeq p_1 e^{-\Gamma_{S} t} + p_2 e^{-\Gamma t} + p_3 e^{-\Gamma_{s} t}$, having super-radiant
$\Gamma_S > \Gamma$, single-atom $\Gamma$, and sub-radiant $\Gamma_s < \Gamma$ decay rates. }
\label{fig:gamma2LAs}
\end{figure}

\begin{figure}[t]
\includegraphics[width=1.0\linewidth]{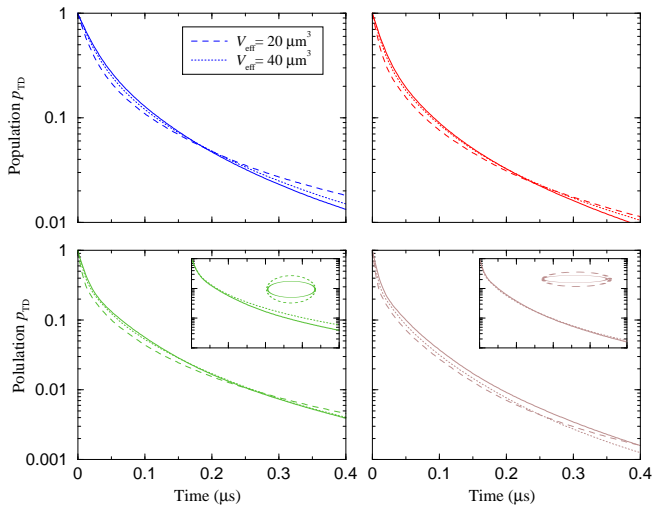}
\caption{Same as in Fig.~\ref{fig:gamma2LAs} (thin solid lines are for reference,
with the upper-left, upper-right, lower-left and lower-right panels corresponding
to decreasing $\sigma_{x,y}$ and increasing $\sigma_z$), but for $N = 2000$ atoms,
in traps with the same volume $V_{\mathrm{eff}} = (2 \pi)^{3/2} 8 \,\mu\mathrm{m}^3$ 
but twice the atom density (dashed lines), or in traps with rescaled dimensions 
$\sigma_{x,y,z} \to \sqrt[3]{2}\sigma_{x,y,z}$ and twice the volume 
$V_{\mathrm{eff}}  = (2 \pi)^{3/2} 16 \,\mu\mathrm{m}^3$ but the same density 
as in Fig.~\ref{fig:gamma2LAs} (dotted lines).
The insets in the lower panels show the decay dynamics for $N = 2000$ atoms 
in the elongated traps with the same lengths $\sigma_{z} = 4,8\,\mu\mathrm{m}$ 
but larger widths $\sigma_{x,y} \to \sqrt{2}\sigma_{x,y}$ and twice  the volume
$V_{\mathrm{eff}}  = (2 \pi)^{3/2} 16 \,\mu\mathrm{m}^3$, i.e. the same density as in Fig.~\ref{fig:gamma2LAs}. }
\label{fig:gamma2LAsN2k}
\end{figure}

Let us assume that at some initial time $t=0$ the atoms are prepared by a laser in the collective single-excitation
(timed-Dicke \cite{Scully2006}) state
\begin{equation}
\ket{E_{\mr{TD}}} = \frac{1}{\sqrt{N}} \sum_{j=1}^N e^{i \bm{k}_c \cdot \bm{r}_j} \ket{e_j} 
= \frac{1}{\sqrt{N}} \sum_{j=1}^N \ket{\tilde{e}_j}, \label{eq:TDstate}
\end{equation}
with $\ket{\tilde{e}_j} \equiv e^{i \bm{k}_c \cdot \bm{r}_j} \ket{e_j}$ and $\bm{k}_c \parallel \hat{\bm{z}}$.
We expand the state of the atomic ensemble as
$\ket{\Psi} = \sum_j b_j e^{-i \omega_e t} \ket{e_j} \equiv \sum_j \tilde{b}_j e^{-i \omega_e t} \ket{\tilde{e}_j}$,
where the slowly varying in time and space excited state amplitudes $\tilde{b}_j = e^{-i \bm{k}_c \cdot \bm{r}_j} b_j$
obey the equations
\begin{equation}
\partial_t \tilde{b}_j = - \frac{1}{2} \Gamma \, \tilde{b}_j
- \frac{1}{2} \Gamma \sum_{i\neq j} F_{ji} e^{i \bm{k}_c \cdot \bm{r}_{ij}} \tilde{b}_i  \label{eqs:tildebjGF}
\end{equation}
with the initial conditions $\tilde{b}_j (0) = \frac{1}{\sqrt{N}} \; \forall \;j$.
For a non-interacting atomic ensemble, $F_{ji} \to 0$, 
i.e., in the dilute regime of large mean interatomic separation $\expv{r_{ij}} \gtrsim N^{1/4} (\sigma_z \lambda_e)^{1/2}$, 
the initial state $\ket{\Psi(0)} = \ket{E_{\mr{TD}}}$ will decay with the single-atom rate $\Gamma$ to the collective
ground state $\ket{G}$ and emit a photon with the spatial profile of Eq.~\eqref{eq:Erffapprx}.
But in the high-density regime, the interatomic dipole-dipole interaction mediated by the multiple scattering
of the photon by the atoms, significantly modifies this behavior, resulting in both accelerated (super-radiant)
decay with rate $\Gamma_S > \Gamma$ and decelerated (sub-radiant) decay with the rate $\Gamma_s < \Gamma$,
as seen in Fig.~\ref{fig:gamma2LAs}. Moreover, for a fixed mean density of the atom cloud, the super- and sub-radiant
decays strongly depend on the geometry of the atom cloud: atoms in an elongated trap, $\sigma_{z} > \sigma_{x,y}$,
typically decay faster, which can be attributed to the constructive interference of the photon emission
(scattering) in the forward direction with larger optical depth.
We can approximate the super-radiant decay rate as
\begin{equation}
\Gamma_S \approx \mc{G} \frac{N}{k_e^2 \sigma_{x,y}^2} \Gamma , \label{eq:GammaS}
\end{equation}
where the numerical factor $\mc{G}$ depends on the geometry of the atom cloud
($\mc{G} \approx \frac{4}{3}, \frac{5}{6}, \frac{2}{3}, \frac{2}{5}$ for the four geometries
shown in Figs.~\ref{fig:gamma2LAs}, \ref{fig:gamma2LAsN2k}).
This is consistent with the previously derived results for isotropic dipoles or lower atom densities    
\cite{Mazetz2007,SvidzinskyScully2008,Svidzinsky2008,Svidzinsky2010,Bienaime2012,Araujo2016,Roof2016,Kuraptsev2017,Araujo2018,Maximo2020},
since the interatomic interactions are predominantly long-range, $F_{ji} \simeq \frac{e^{i k_e r_{ij}}}{i k_e r_{ij}}$, 
and the contribution of the near-field terms $ \propto (k_e r_{ij})^{-2(3)}$ is small when $k_e \expv{r_{ij}} \gg 1$.

Increasing the atom number $N$ and thereby the density in the same trapping volume accelerates the super-radiant decay,
i.e. further increases $\Gamma_S$, and decelerates the sub-radiant decay, i.e. further decreases $\Gamma_s$,
of the collective timed-Dicke state as seen in Fig.~\ref{fig:gamma2LAsN2k}. On the other hand, increasing
the atom number and proportionally the trapping volume to keep the atom density constant, we observe smaller
modification of the super-radiant decay, consistent with Eq.~(\ref{eq:GammaS}).
Finally, for a fixed atom density, changing only the width of the trap, but not its length, increases the
sub-radiant fraction of the initial population (see the insets of Fig.~\ref{fig:gamma2LAsN2k}), which indicates
that the sub-radiant dynamics is mostly governed by the multiple scattering of the photons off the $z$ axis,
while the super-radiant emission happens mostly in the forward direction along $z$.

\subsection{Single-excitation spectrum of the atoms}
\label{sec:tla1espct}

Equation (\ref{eqs:tildebjGF}) implies an effective non-Hermitian Hamiltonian for $N$ interacting atoms:
\begin{eqnarray}
H_{\mr{eff}} &=& \sum_{j=1}^N \hbar \left( \omega_{e} - i \frac{\Gamma}{2} \right) \ket{\tilde{e}_j}\bra{\tilde{e}_j}
\nonumber \\ & &
- i \frac{\Gamma}{2} \sum_{j=1}^N  \sum_{j' \neq j}^N F_{jj'} e^{i \bm{k}_c \cdot \bm{r}_{j'j}} \ket{\tilde{e}_{j'}}\bra{\tilde{e}_j} .
\label{eq:Heff}
\end{eqnarray}
The solution of the eigenvalue problem $H_{\mr{eff}} \ket{\Psi} = \hbar \lambda \ket{\Psi}$ results in $N$
generally non-orthogonal (right) eigenstates $\ket{\Psi_n}$ with complex eigenvalues $\lambda_n$.
The real part of each eigenvalue $\mr{Re} (\lambda_n) = \om_e + \delta_n$ determines the level shift $\delta_n$
of the corresponding eigenstate from the single-atom resonance $\omega_e$, while the imaginary part
$\mr{Im} (\lambda_n) = - \gamma_n$ yields the level width or (half-)decay rate $\gamma_n$ of the eigenstate.

\begin{figure}[t]
\includegraphics[width=1.0\linewidth]{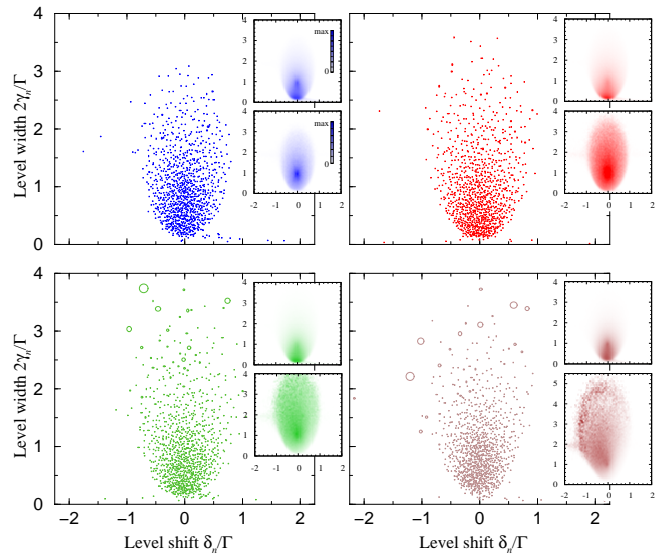}
\caption{Eigenvalues of the effective Hamiltonian (\ref{eq:Heff}) for $N = 1000$ atoms in four 
different traps as in Fig.~\ref{fig:gamma2LAs} (same color code, with the upper-left, upper-right, 
lower-left and lower-right panels corresponding to decreasing $\sigma_{x,y}$ and increasing $\sigma_z$). 
Main panels show the eigenvalues, as obtained for a single realization 
of the ensemble of atoms at random positions in the trapping volume; each eigenvalue
$\lambda_n = (\omega_e + \delta_n) - i \gamma_n$ is shown as a circle centered at the corresponding
$[\delta_n, \gamma_n]$ with the radius equal to the norm $|\braket{E_{\mr{TD}}}{\Psi_n}|^2$ of the Franck-Condon
overlap of the eigenstate $\ket{\Psi_n}$ with the single-excitation state $\ket{E_{\mr{TD}}}$ of Eq.~(\ref{eq:TDstate}).
The upper inset in each panel shows the spectrum of eigenvalues averaged over $10^3$ random realizations
of the ensemble, while the lower inset shows the same spectrum with each eigenvalue weighted by the
corresponding FC factor (the shading is in arbitrary units, for best visibility).}
\label{fig:ReImLasAv}
\end{figure}

Note that for a non-interacting system with $F_{jj'} = 0 \; \forall \; j' \neq j$, all $N$ eigenstates would
be degenerate, $\lambda_n = \omega_e - i \Gamma/2$, and we could construct one ``bright'' eigenstate
$\ket{\Psi_B} = \ket{E_{\mr{TD}}}$ that corresponds to the timed-Dicke state of Eq.~(\ref{eq:TDstate}),
while all the other eigenstates would be ``dark'', $\braket{E_{\mr{TD}}}{\Psi_{n \neq B}} = 0$, i.e., not accessible
from either the ground or the storage state by a uniform laser field with wavevector $\bm{k}_c$ (see below).

\begin{figure}[t]
\includegraphics[width=1.0\linewidth]{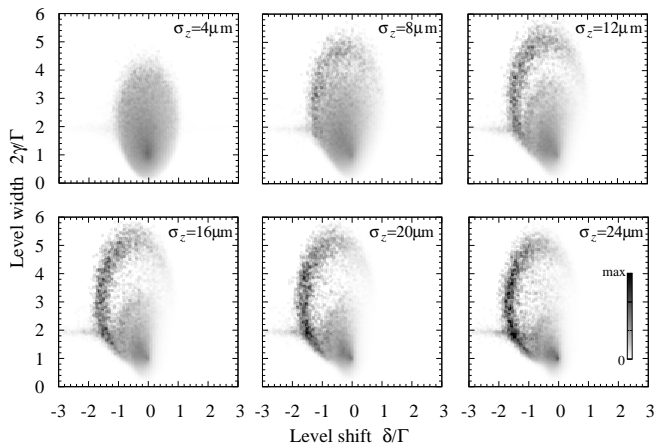}
\caption{Density of eigenvalues weighted by the corresponding FC factors $|\braket{E_{\mr{TD}}}{\Psi_n}|^2$
averaged over $10^3$ random realizations of the ensemble of $N = 1000$ atoms in the elongated harmonic
traps with lengths $\sigma_z=4,8, \ldots, 24 \,\mu$m and widths $\sigma_{x,y} = \sqrt{8/\sigma_z}$,
i.e., the same effective volume $V_{\mathrm{eff}} = (2 \pi)^{3/2} \sigma_x \sigma_y \sigma_z = (2 \pi)^{3/2} 8\,\mu\mathrm{m}^3$,
and mean interatomic separation $\expv{|\bm{r}_{ij}|} = \sqrt[3]{V_{\mathrm{eff}}/N} \simeq 0.5\,\mu\mathrm{m}$
as in Figs.~\ref{fig:gamma2LAs} and \ref{fig:ReImLasAv}.}
\label{fig:eigsFClong}
\end{figure}

In Fig.~\ref{fig:ReImLasAv} we show the spectrum of the effective Hamiltonian (\ref{eq:Heff}) for each of the
four geometries of the trap with $N=1000$ atoms. Since the eigenstates $\ket{\Psi_n}$ of the interacting
system can be populated either from the ground state $\ket{G}$ or from the collective storage state $\ket{S}$
using a near-resonant laser with wavevector $\bm{k}_c$, we calculate the Franck-Condon (FC) overlap
$\braket{E_{\mr{TD}}}{\Psi_n}$ of each eigenstate with the timed-Dicke state of Eq.~(\ref{eq:TDstate}).
We observe that the spectrum of the effective Hamiltonian has super-radiant, $\gamma_n > \Gamma/2$,
and sub-radiant, $\gamma_n < \Gamma/2$, states, and most of the sub-radiant states have small
level shifts $|\delta_n| \lesssim \Gamma$, while the super-radiant states have a broader spectrum
of shifts from the atomic transition resonance $\omega_e$. For trap dimensions $\sigma_{x,y} \sim \sigma_z$
the averaged spectrum, and the spectrum of eigenstates weighted by the FC factors, are approximately
symmetric about the resonance, $\delta=0$. But in the elongated trap  $\sigma_{z} \gg \sigma_{x,y}$
the super-radiant states with the largest FC factors tend to be shifted towards the lower frequencies
$\delta < 0$ (see the lower right panel of Fig.~\ref{fig:ReImLasAv}).
This effect is even better pronounced for highly elongated traps, as shown in
Fig.~\ref{fig:eigsFClong}, and it is closely related to the collective shift of resonant light
scattering by a one-dimensional atomic medium, due to constructive interference of the red-detuned light,
as reported in \cite{Glicenstein2020}.

\begin{figure}[t]
\includegraphics[width=1.0\linewidth]{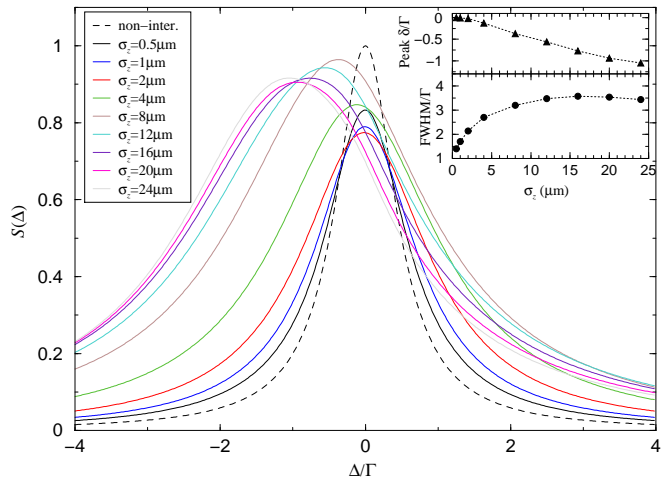}
\caption{Excitation spectrum $S(\Delta)$ of Eq.~(\ref{eq:spectS}) for $N=1000$ atoms in harmonic traps with
different length $\sigma_z$ and width $\sigma_{x,y} = \sqrt{8/\sigma_z}$, i.e., the same effective volume
$V_{\mathrm{eff}} = (2 \pi)^{3/2} \sigma_x \sigma_y \sigma_z = (2 \pi)^{3/2} 8\,\mu\mathrm{m}^3$.
The spectrum is progressively broadened and shifted towards the negative detuning $\Delta$ with increasing $\sigma_z$. 
For each geometry, the shown spectrum is averaged over $10^3$ random realizations of the atomic ensemble.
Insets show the peak position and full width at half maximum (FWHM) of each spectrum.}
\label{fig:spect}
\end{figure}

The amplitude $B_n(t)$ of an eigenstate $\ket{\Psi_n}$ excited with probability $|B_n(0)|^2$ at
time $t=0$ evolves according to $B_n(t) = B_n(0) e^{-i \lambda_n t}$. Taking the Fourier transform
$\int_0^{\infty} dt B_n(t) e^{i \omega t}$, we can then associate with each eigenstate having the decay
rate $\gamma_n$ and level shift $\delta_n$ a Lorentzian emission/absorption line
$\frac{|B_n(0)|^2 \gamma_n^2}{(\Delta - \delta_n)^2 + \gamma_n^2}$ with $\Delta = \omega - \omega_e$.
Since the excitation probability of each eigenstate from either the ground or the storage state
via a laser with wavevector $\bm{k}_c$ is proportional to $|\braket{E_{\mr{TD}}}{\Psi_n}|^2$, we
can then define the excitation (absorption) spectrum of the systems as
\begin{equation}
S(\Delta) = \sum_{n=1}^N \frac{|\braket{E_{\mr{TD}}}{\Psi_n}|^2 \gamma_n^2}{(\Delta - \delta_n)^2 + \gamma_n^2} . 
\label{eq:spectS}
\end{equation}
In Fig.~\ref{fig:spect} we plot $S(\Delta)$ for various geometries of the atomic ensemble. With increasing
length of the atomic ensemble, the excitation spectrum is progressively broadened and shifted
towards the lower frequencies, i.e. negative detuning $\Delta$. This is expected from Fig.~\ref{fig:eigsFClong},
which demonstrates that in the highly elongated atomic ensembles the eigenstates with the largest FC factors 
are super-radiant ($\gamma > \Gamma/2$) and red-shifted ($\delta < 0$) with respect to the single-atom resonance $\omega_e$.
The red shift of the collective resonance in elongated atomic ensembles has been experimentally observed 
in Ref.~\cite{Roof2016}.

\subsection{Angular emission profile}

\begin{figure*}
\includegraphics[width=1.0\linewidth]{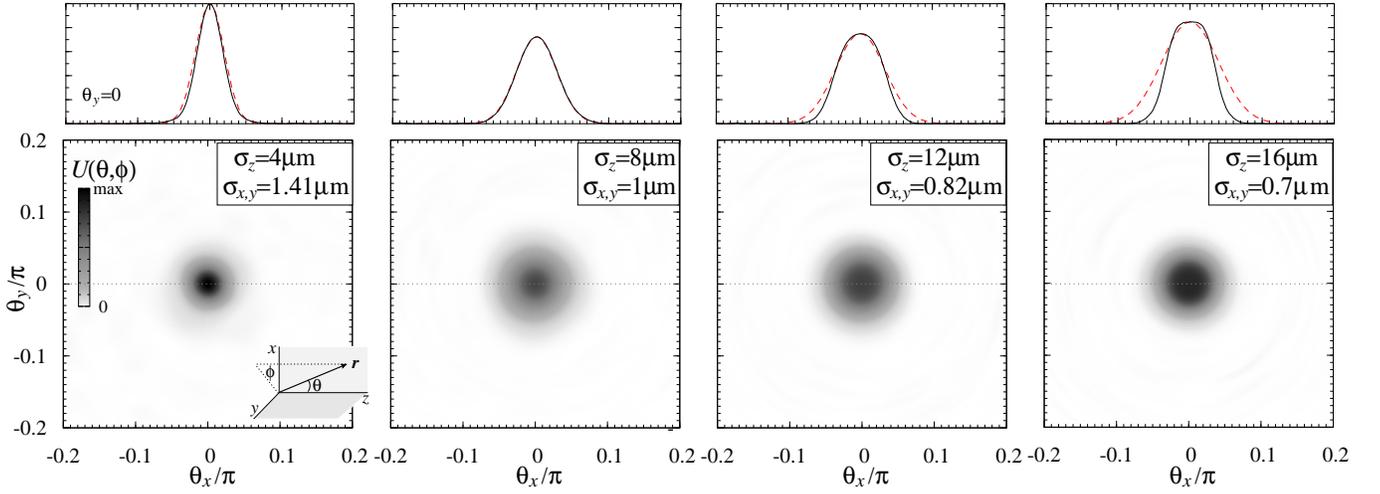}
\caption{Angular probability distribution $U(\theta,\phi)$ of the photon emitted
in the $z$ direction, as a function of $\theta_x = \theta \cos(\phi)$ and $\theta_y = \theta \sin(\phi)$
with $\theta$ the polar and $\phi$ the azimuthal angles, for $N = 1000$ atoms in elongated harmonic
traps with the same effective volume $V_{\mathrm{eff}} = (2 \pi)^{3/2} \sigma_x \sigma_y \sigma_z$ and
different aspect ratios $\sigma_{x,y}/\sigma_z$.
The black solid line in the upper inset of each density plot shows $U(\theta_x,\theta_y = 0)$ while the red dashed line
is the Gaussian of Eq.~(\ref{es:Stheta}) with the corresponding width $\Delta \theta = \frac{\sqrt{2}}{k_e \sigma_{x,y}}$.
Only the interval of $|\theta| \leq 0.2 \pi$ is shown as in the remaining solid angle only a weak noisy signal is present.}
\label{fig:Sr}
\end{figure*}

The (far) field emitted by the atoms, Eq.~(\ref{eq:Erff}), 
in terms of the temporally and spatially slowly varying atomic amplitudes $\tilde{b}_j = e^{-i \bm{k}_c \cdot \bm{r}_j} b_j$,
is given by
\begin{eqnarray}
E_{\sigma}^{(\mathrm{ff})}(\bm{r},t) &=& (\hat{\bm{\veps}}_{\bm{r},\sigma} \cdot \hat{\bm{\wp}})
\frac{\wp_{eg} k_e^2}{4\pi \eps_0}  \frac{e^{i (k_er - \om_e t)}}{r} 
\nonumber \\ & & \quad \times
\sum_j  \tilde{b}_j(t - r/c) e^{i (\bm{k}_c - \bm{k}_e) \cdot \bm{r}_j} , \label{eq:Erfftildbj}
\end{eqnarray}
which clearly reveals the phase matching condition $\bm{k}_e \simeq \bm{k}_c \parallel \hat{\bm{z}}$
for constructive interference of photon emission. The intensity of the emitted radiation in
the direction of $\bm{r}$ is $I_{\sigma}(\bm{r},t) = \frac{\eps_0 c}{2} |E_{\sigma} (\bm{r},t)|^2$,
while the total radiation (energy) collected by an ideal detector at position $\bm{r}$ is
$U(\bm{r})\, \delta s = \sum_{\sigma=1,2} \int_0^{\infty} dt I_{\sigma}(\bm{r},t) \, \delta s$, where
$\delta s$ is the surface element, or detector cross-section (pixel size), in the plane perpendicular
to $\bm{r}$. From the discussion of Eq.~(\ref{eq:Erffapprx}), we expect that the angular distribution
of the radiation emitted into the phase-matched direction $z$ can be approximated by a Gaussian
\begin{equation}
U (\theta) \propto e^{-2 \theta^2/\Delta \theta^2}  \label{es:Stheta}
\end{equation}
with the beam divergence $\Delta \theta = \frac{\lambda_e}{\pi w_0} = \frac{\sqrt{2}}{k_e \sigma_{\perp}}$.
In Fig.~\ref{fig:Sr} we show the angular probability distribution of $U (\theta,\phi)$ which is highly
peaked around $\theta =0$ due to cooperative photon emission into the phase-matched direction, while
for larger angles $\theta > \Delta \theta$ we observe a weak background noise due to spontaneous,
uncorrelated photon emission by atoms at random positions. For large enough width $\sigma_{\perp}$ of
the atomic cloud, the emitted radiation profile is indeed Gaussian with the angular width $\Delta \theta$.
But as the transverse width of the cloud becomes comparable to, or smaller than, the wavelength,
$\sigma_{\perp} \lesssim \lambda_e = 0.780\,\mu\mr{m}$, the angular profile of the beam starts
to strongly deviate from the Gaussian, i.e., it becomes narrower than the corresponding $\Delta \theta$
and develops a ``flat top''. We have checked that the narrowing effect is also present in the ensemble
of non-interacting atoms, but the flattening of the top is effected by interatomic interactions.
Thus, to maximize the collection of radiation from a highly elongated atomic cloud, one should
engineer a lens with an appropriate non-circular curvature.

\begin{figure}[b]
\includegraphics[width=0.9\linewidth]{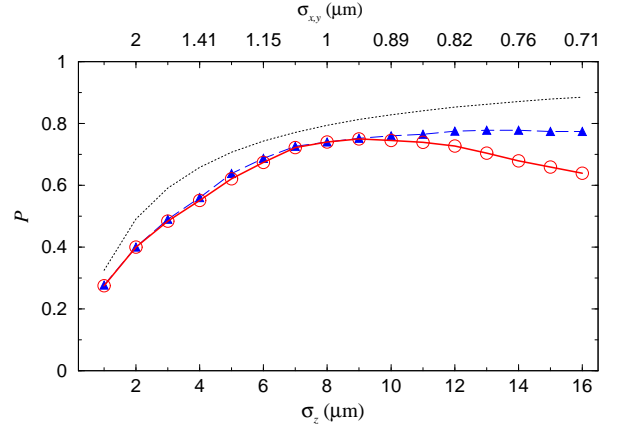}
\caption{Probability $P$ of Eq.~(\ref{eq:Pemitint}) for collecting the emitted photon into the Gaussian mode
of waist $w_0 = \sqrt{2} \sigma_{x,y}$ as a function of length $\sigma_z$ (lower horizontal axis) or width
$\sigma_{x,y}$ (upper horizontal axis) of a cloud of $N = 1000$ interacting atoms with fixed effective volume
$V_{\mathrm{eff}} = (2 \pi)^{3/2} \sigma_x \sigma_y \sigma_z = (2 \pi)^{3/2} 8$ (red solid line with open circles).
Also shown is the total probability of photon emission into the $z$ direction within the solid angle
$\Omega_f = \pi (2 \, \Delta \theta)^2$ (blue dashed line with filled triangles).
For comparison, we also show the approximate analytic results of Eq.~(\ref{eq:PDeOm-nia}) for an ensemble
of non-interacting atoms (black dotted line). }
\label{fig:Pemit}
\end{figure}

Our aim is to determine the probability of collecting the photon by an appropriate paraxial optics
into the Gaussian mode
\begin{equation}
\phi_{k_e}(\bm{r},t) = \frac{\zeta_{k_e} e^{-i ck_e t}}{q_{k_e}^*(z)}
\exp \left[ i k_e \left( z + \frac{x^2+y^2}{2q_{k_e}^*(z)} \right) \right]
\end{equation}
with the wavevector $k_e$ and waist $w_0 = \sqrt{2} \sigma_{x,y}$. To this end, we calculate
the overlap of the far field $E_{\sigma}^{(\mr{ff})}$ with $\phi_{k_e}$ on a spherical surface with
large radius $r = \sqrt{x^2+y^2+z^2} \gg \sigma_{x,y,z},\zeta_{k_e}$ integrating over the $4\pi$ solid angle,
$\int_{4\pi} d \Omega_r [ E_{\sigma}^{(\mr{ff})} (\bm{r},t) \, \phi_{k_e}^*(\bm{r},t)]$.
The probability for the emitted photon to be collected into the Gaussian mode $\phi_{k_e}^*(\bm{r})$ is then
\begin{equation}
P = \frac{ \sum_{\sigma=1,2} \left| \int_{4\pi} d \Omega_r \,
\int_0^{\infty} dt  \frac{\eps_0 c}{2} [E_{\sigma}^{(\mr{ff})} (\bm{r},t) \, \phi_{k_e}^*(\bm{r},t)] \right|^2}
{\int d \Omega_r U(\bm{r}) \int d \Omega_r |\phi_{k_e} (\bm{r}) |^2} . \label{eq:Pemitint}
\end{equation}
In Fig.~\ref{fig:Pemit} we show $P$ for various lengths $\sigma_z$ and the corresponding
widths $\sigma_{x,y}$ or the atomic ensemble with the same effective volume and density.
Note that the portion of the radiation emitted into the phase-matched direction $z$ grows monotonically with
increasing cloud length $\sigma_z$, although this growth nearly stops once the decreasing transverse width
of the cloud becomes comparable to the wavelength, $\sigma_{x,y} \lesssim \lambda_e = 0.780\,\mu\mr{m}$.
The probability $P$ of the photon to be emitted into the appropriate Gaussian mode also grows initially with
increasing $\sigma_z$, but it is peaked around $\sigma_z \simeq 9 \,\mu\mr{m}$ ($\sigma_{x,y} \lesssim 1 \,\mu\mr{m}$)
and then decreases, since the spatial profile of the emitted radiation increasingly deviates from the Gaussian
for narrower atom clouds [cf. upper insets in Fig.~\ref{fig:Sr}].
Note finally that the approximate analytic result of Eq.~(\ref{eq:PDeOm-nia}) for non-interacting atomic ensemble
predicts larger emission probabilities into the Gaussian mode, but this result also does not take into account
the narrowing of the emission profile for highly elongated atom clouds.

\section{Three-level atomic medium}
\label{sec:threeLAs}

In the previous section, we considered an ensemble of two-level atoms and assumed that initially 
the system is somehow prepared in the ideal timed-Dicke state with single collective excitation.
But starting from the collective ground or storage state of the atomic ensemble, the preparation
of the timed-Dicke state may be hindered by the strong interatomic interactions leading to spectral
broadening and suppression of the transition to the collective excited states, as discussed in Sec. \ref{sec:tla1espct}. 
We therefore consider now all three atomic levels, and assume that the initially populated collective 
storage state $\ket{S} = \frac{1}{\sqrt{N}} \sum_{j=1}^N \ket{s_j}$ is coupled to the excited state by a laser
with time-dependent Rabi frequency $\Omega(t)$ and detuning $\Delta_c$, as shown in Fig.~\ref{fig:Scheme}.
The atomic amplitudes obey the equations
\begin{subequations}
\label{eqs:ncjtildebj}
\begin{eqnarray}
\partial_t c_j &=& i \Omega^* \tilde{b}_j e^{i \Delta_c t}, \label{eqs:ncj} \\
\partial_t \tilde{b}_j &=& i \Omega  c_j e^{- i \Delta_c t}  - \frac{1}{2} \Gamma \, \tilde{b}_j
- \frac{1}{2} \Gamma \sum_{i\neq j} F_{ji} e^{i \bm{k}_c \cdot \bm{r}_{ij}} \tilde{b}_i . \qquad \label{eqs:tildebj}
\end{eqnarray}
\end{subequations}

For non-interacting atoms \cite{Kurko}, $F_{ji} = 0$, assuming a resonant laser $\Delta_c = 0$ 
with sufficiently weak Rabi frequency $|\Omega| < \Gamma$, we can set $\partial_t \tilde{b}_j = 0$, 
obtaining $\tilde{b}_j \simeq i \frac{\Omega}{\Gamma/2}  e^{i \bm{k}_c \cdot \bm{r}_j}c_j$. 
Substituting this into Eq.~(\ref{eqs:ncj}) and performing the integration, we have
\begin{subequations}
\label{eq:cbjt}
\begin{eqnarray}
c_j (t) &\simeq & c_j (0) \exp \left[ - \int_0^t dt' \frac{|\Omega(t')|^2}{\Gamma/2} \right] , \\
\tilde{b}_j(t) &\simeq & i \frac{\Omega(t)}{\Gamma/2} c_j (t)  , 
\end{eqnarray}
\end{subequations}
with the initial condition $c_j(0) = 1/\sqrt{N} \; \forall \; j \in [1,N]$. Using this solution
in Eq.~(\ref{eq:Erff}) or Eq.~(\ref{eq:Erfftildbj}), we obtain
\begin{subequations}
\begin{eqnarray}
E^{(\mathrm{ff})}_{\sigma}(\bm{r},t) &=&
i \frac{\wp_{eg} k_e^2}{4\pi \eps_0} \frac{\beta(t-r/c)}{\sqrt{N}} e^{- i \om_e t} \, \mc{E}_{\sigma}(\bm{r}), \\
\mc{E}_{\sigma}(\bm{r}) &\equiv & (\hat{\bm{\veps}}_{\bm{r},\sigma} \cdot \hat{\bm{\wp}})
\frac{e^{i k_e r }}{r} \sum_j e^{i (\bm{k}_c - \bm{k}_e) \cdot \bm{r}_j} ,
\end{eqnarray}
\end{subequations}
which, apart from the time dependence contained in
$\beta(t) \equiv  \frac{\Omega(t)}{\Gamma/2} \exp \left[ - \int_0^t dt' \frac{|\Omega(t')|^2}{\Gamma/2} \right] $
and field polarization, is the same as Eqs.~(\ref{eqs:Erffini}) with all the consequences discussed there.
Detailed treatment of Raman conversion of collective atomic excitation in a non-interacting ensemble to a photon 
and its optimal collection via paraxial optics is presented in \cite{Kurko}.

\subsection{Dynamics of population transfer}

\begin{figure}[t]
\includegraphics[width=1.0\linewidth]{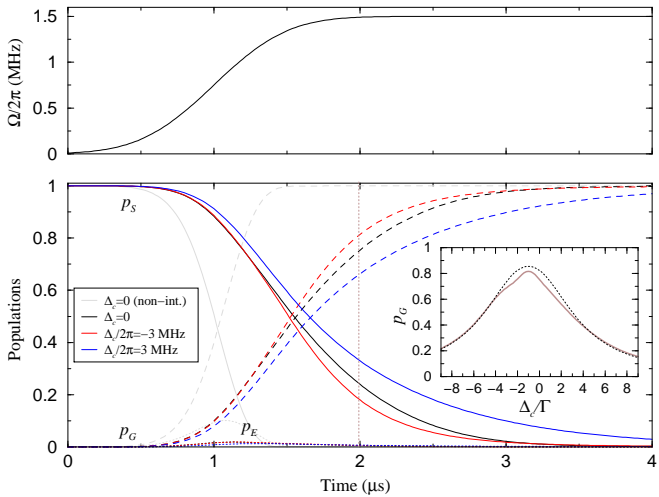}
\caption{Dynamics of population transfer of the atoms from the storage state to the excited state
that decays to the ground state with the emission of a photon.
The upper panel shows the time-dependence of the coupling field Rabi frequency
$\Omega(t) = \Omega_0 \frac{1}{2} \big[ 1 + \mr{erf}(\frac{t-t_0}{\sqrt{2}\sigma_t}) \big]$
with $t_0 = 1\,\mu\mr{s}$ and $\sigma_t = 0.4\,\mu\mr{s}$.
The lower panel shows the populations of the storage $p_S$, excited $p_E$ and ground $p_G$ states
for three different detunings $\Delta_c$ of the coupling field 
(with the detunings $\Delta_c/2\pi = \pm 3\:$MHz leading to slower/faster transfer 
compared to the $\Delta_c = 0$ case), as obtained from averaging over $10^2$ random 
realizations of the ensemble for $N=1000$ atoms in a harmonic trap with dimensions
$\sigma_z = 8\,\mu\mr{m}$ and $\sigma_{x,y} = 1\,\mu\mr{m}$. Also shown are the populations
for an ensemble of non-interacting atoms under the same driving and decay conditions (light-gray curves).
The inset shows the transfer probability to the ground state $p_G$ at time $t=2\,\mu\mr{s}$
as a function of detunings $\Delta_c$, as obtained from a single realization of the random atomic ensemble
(thick solid brown line), and as obtained from the approximate analytic solution of Eqs.~(\ref{eqs:cbSEG})
with parameters $\Omega_{\mr{eff}} = 0.92 \Omega$, $\delta_E = - 1.0 \Gamma$ and $\Gamma_S = 6.0 \Gamma$
(thin dotted black line); note that this value for $\Gamma_S$ was obtained in Fig.~\ref{fig:gamma2LAs} by
fitting the decay curve for an initially excited atomic ensemble.}
\label{fig:stRamSpct}
\end{figure}

For strongly interacting atoms, the above simple solution does not apply, and we resort
to the numerical solutions of the atomic equations of motion (\ref{eqs:ncjtildebj}).
In Fig.~\ref{fig:stRamSpct} we show the dynamics of populations of the storage state $p_S = \sum_j |c_j|^2$,
the excited state $p_E = \sum_j |\tilde{b}_j|^2$ and the ground state $p_G = 1 - p_S - p_E$
upon applying to the ensemble a coupling field with a smooth time-dependent Rabi frequency
$\Omega(t)$ and various detunings $\Delta_c$ from the unperturbed atomic transition $\ket{s} \to \ket{e}$.
It follows from the above discussion that the transition from the symmetric storage state $\ket{S}$ to the
collective excited state $\ket{E_{\mr{TD}}}$ driven by a uniform laser with wavevector $\bm{k}_c \parallel \hat{\bm{z}}$
is suppressed by either the small FC factors $\braket{E_{\mr{TD}}}{\Psi_n}$ or large widths $\gamma_n$ of the single
excitation eigenstates $\ket{\Psi_n}$, which results in much slower population transfer as compared to the non-interacting atoms.
Moreover, since the spectrum of the eigenstates $\ket{\Psi_n}$ weighted by the corresponding FC factors
is asymmetric and red-shifted from the atomic resonance frequency $\omega_e$, we observe stronger excitation,
followed by decay, for negative detuning $\Delta_c < 0$. The inset in Fig.~\ref{fig:stRamSpct} shows the transfer
probability to $p_G$ at an intermediate time $t=2\,\mu\mr{s}$ as a function of $\Delta_c$, which is closely
related to the excitation spectrum $S(\Delta)$ of Fig.~\ref{fig:spect}.

It is instructive to consider an effective three-level system with the ground state $\ket{G}$,
the storage state $\ket{S}$ and an excited state $\ket{E}$ which is shifted from the single atom
resonance by $\delta_E$ and decays to the ground state with rate $\Gamma_S > \Gamma$.
The initially populated storage state is coupled to the excited state $\ket{E}$ with
an effective Rabi frequency $\Omega_{\mr{eff}}$.
The amplitudes $c$ and $b$ of the storage and excited states obey the equations
\begin{subequations}
\label{eqs:cbSEG}
\begin{eqnarray}
\partial_t c &=& i \Omega^*_{\mr{eff}} b \, e^{i (\Delta_c -\delta_E)  t}, \label{eqs:cS} \\
\partial_t b &=& i  \Omega_{\mr{eff}}   c \, e^{- i (\Delta_c -\delta_E) t}  - \frac{1}{2} \Gamma_S b ,
\end{eqnarray}
\end{subequations}
which have an approximate analytic solution similar to that of Eqs.~(\ref{eq:cbjt}), namely
\begin{subequations}
\begin{eqnarray}
c(t) &\simeq & c(0) \exp\left[ - \int_0^t dt' \frac{|\Omega_{\mr{eff}}(t')|^2}{\Gamma_S/2 - i (\Delta_c -\delta_E)} \right] , \\
b(t) &\simeq & i \frac{\Omega_{\mr{eff}}(t) e^{- i (\Delta_c -\delta_E) t}}{\Gamma_S/2 - i (\Delta_c -\delta_E)} \, c(t),
\end{eqnarray}
\end{subequations}
with $c(0) = 1$.
The probability of population transfer to the ground state can then be approximated as $p_G(t) = 1 - |c(t)|^2 - |b(t)|^2$.
In the inset in Fig.~\ref{fig:stRamSpct} we compare this analytic result with the exact numerical result and find
reasonable agreement for appropriate parameters $\Omega_{\mr{eff}}$, $\delta_E$ and $\Gamma_S$.

\subsection{Radiation field}

\begin{figure}[t]
\includegraphics[width=1.0\linewidth]{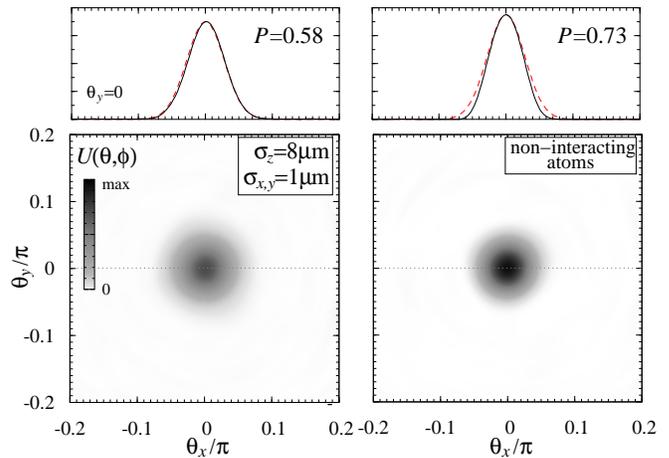}
\caption{Angular probability distribution $U(\theta,\phi)$ of the photon emitted in the $z$ direction
for Raman excitation of the atomic cloud with all the parameters the same as in Fig.~\ref{fig:stRamSpct}.
The left panel shows $U(\theta,\phi)$ vs. $\theta_x = \theta \cos(\phi)$ and $\theta_y = \theta \sin(\phi)$
as obtained for a single realization of the random atomic ensemble;
For large integration times, the photon emission pattern is the same for different detunings $\Delta_c$
of the coupling field, and the probability of cooperative emission into a Gaussian mode of waist
$w_0 = \sqrt{2} \sigma_{x,y}$ is $P \simeq 0.583$, with the remaining radiation incoherently scattered
into all $4\pi$ directions.
The right panel shows $U(\theta,\phi)$ for a non-interacting atomic ensemble under the otherwise identical
conditions, leading to $P \simeq 0.727$ [for comparison, the analytic result of Eq.~(\ref{eq:PDeOm-nia})
and Fig.~\ref{fig:Pemit} is $P= 0.794$].
The black solid line in the upper inset of each density plot shows $U(\theta_x,\theta_y=0)$
while the red dashed line is the Gaussian of Eq.~(\ref{es:Stheta}) with the corresponding
width $\Delta \theta = \frac{\sqrt{2}}{k_e \sigma_{x,y}}$.}
\label{fig:SrRaman}
\end{figure}

The angular distribution $U(\theta,\phi)$ of the emitted radiation is shown in Fig.~\ref{fig:SrRaman}.
We observe that for an interacting atomic ensemble the spatial profile of the radiation emitted in the phase-matched
direction $z$ closely matches a Gaussian mode of waist $w_0 = \sqrt{2} \sigma_{x,y}$. But the probability
of cooperative emission into this Gaussian mode is rather small, $P \simeq 0.58$, since multiple photon scattering
by the atoms results in large fraction $1-P \gtrsim 0.4$ of the radiation to be incoherently emitted into all
$4\pi$ directions. For comparison, for a non-interacting atomic ensemble we obtain a much larger $P \simeq 0.73$,
even though the forward emitted radiation has somewhat narrower angular distribution than that of the expected
Gaussian mode [for a slightly narrower Gaussian collection mode, we obtain $P \simeq 0.77$ close to the theoretical
result of Eq.~(\ref{eq:PDeOm-nia}) and Fig.~\ref{fig:Pemit}]. Hence, dilute atomic ensembles with reduced
multiple photon scattering seem to be better suited for achieving higher efficiency of photon collection into 
appropriate Gaussian modes \cite{Kurko}. 

Remarkably, even though the dynamics of population transfer between the collective atomic states 
depends on the detuning $\Delta_c$ of the coupling field [cf. Fig.~\ref{fig:stRamSpct}],
for large integration times, when the population of the symmetric storage state of the atomic ensemble is completely
depleted, $p_S =0$ and $p_G \simeq 1$, we obtain the same emission pattern $U(\theta,\phi)$ and photon collection
probability $P \simeq 0.58$ of the interacting atomic ensemble for any detuning $\Delta_c$.
Thus the atomic ensemble indeed behaves as an effective three-level medium with a single broad intermediate excited state
[cf. inset in Fig.~\ref{fig:stRamSpct}], rather than as a collection of single-excitation states $\ket{\Psi_n}$
with different widths and coupling strengths.

\section{Conclusions}
\label{sec:concl}

A topic of great interest of current research is the interaction of light with regular arrays of strongly (dipole-dipole) 
interacting atoms \cite{Bettles2016,Facchinetti2016,Asenjo-Garcia2017,Shahmoon2017,Grankin2018,Guimond2019,Rui2020}.
Such systems possess cooperative resonances corresponding to super- and sub-radiant optical modes and can serve
as, e.g., perfect optical mirrors \cite{Bettles2016,Shahmoon2017,Guimond2019,Rui2020} or tailored, highly-efficient
photon emitters into the desired spatial modes \cite{Asenjo-Garcia2017,Grankin2018}. These unique properties, however,
critically depend on the periodic, defect-free spatial arrangement of single atoms in lattices with subwavelength
spacing.

Here, we have considered a high-density regime of random atomic ensembles with the subwavelength mean interatomic distance.
This system permits a much smaller degree of control of the super- and sub-radiance and spatial emission pattern of the radiation,
as compared to the perfectly periodic 1D, 2D or 3D arrays of atoms. Yet, the random atomic ensembles are much easier to realize
experimentally in various trapping geometries, which still allow a certain amount of control of their optical properties,
as we have shown above. In particular, we have found that the phase-matched, super-radiant emission of radiation is strongly
enhanced in elongated atomic ensembles, while multiple scattering of photons off the phase-matching direction is mainly
responsible for the sub-radiant emission. It would be interesting to investigate how the super- and sub-radiant collective
modes can be selectively suppressed or converted on demand into each other, and how to control and further enhance the
directionality of the photon emission using, e.g., spatial and/or temporal modulation of the amplitudes and phases of the atoms
in extended traps, which can be accomplished by spatially varying electric or magnetic fields or ac Stark shifts induced by
off-resonant lasers. Finally, studying multiple excitations in strongly-interacting but random atomic ensembles beyond
the liner optical regime would be an interesting and important problem to tackle via development of effective analytic 
and efficient numerical tools.

\section*{Acknowledgments}
We thank \'Arp\'ad Kurk\'o,  Peter Domokos,  Thomas B\ae kkegaard, and J\'ozsef Fort\'agh for useful discussions.
We acknowledge support by the US ARL-CDQI program through cooperative agreement W911NF-15-2-0061.
D.P. was supported by the EU QuantERA Project PACE-IN (GSRT grant No. T11EPA4-00015) and 
by the Alexander von Humboldt Foundation in the framework of the Research Group Linkage Programme.
K.M. was supported by the Carlsberg Foundation through the Semper Ardens Research Project QCooLEU
and by the Danish National Research Foundation Centre of Excellence for Complex Quantum Systems.

\end{document}